\documentclass[runningheads]{llncs}
\usepackage[T1]{fontenc}
\usepackage{graphicx}
\usepackage{caption}
\usepackage{subcaption}
\usepackage{listings} % additional package
\usepackage{placeins} % additional package
\usepackage{hyperref}
\begin{document}
\title{RAI: Flexible Agent Framework for Embodied AI}
\author{ Kajetan Rachwał \inst{1}\orcidID{0000-0003-1524-7877} \and
Maciej Majek \inst{1}\orcidID{0009-0009-9541-8461} \and
Bartłomiej Boczek \inst{1}\orcidID{0009-0006-9097-9655} \and
Kacper Dąbrowski \inst{1}\orcidID{0009-0001-5661-9801} \and
Paweł Liberadzki \inst{1}\orcidID{0000-0002-4072-7605} \and
Adam Dąbrowski \inst{1}\orcidID{0000-0002-2130-0577} \and
Maria Ganzha \inst{2}\orcidID{0000-0001-7714-4844}
}
\authorrunning{K. Rachwał et al.}
\institute{Robotec.AI, Warsaw, Poland \\ \email{\{name.surname\}@robotec.ai} \and Faculty of Mathematics and Information Science, Warsaw University of Technology, Warsaw, Poland}
\maketitle              % typeset the header of the contribution
\begin{abstract}
With an increase in the capabilities of generative language models, a growing interest in embodied AI has followed.
This contribution introduces RAI -- a framework for creating embodied Multi Agent Systems for robotics.
The proposed framework implements tools for Agents' integration with robotic stacks, Large Language Models, and simulations.
It provides out-of-the-box integration with state-of-the-art systems like ROS~2.
It also comes with dedicated mechanisms for the embodiment of Agents.
These mechanisms have been tested on a physical robot,  Husarion ROSBot XL, which was coupled with its digital twin, for rapid prototyping.
Furthermore, these mechanisms have been deployed in two simulations: (1)~robot arm manipulator and (2)~tractor controller.
All of these deployments have been evaluated in terms of their control capabilities, effectiveness of embodiment, and perception ability.
The proposed framework has been used successfully to build systems with multiple agents.
It has demonstrated effectiveness in all the aforementioned tasks.
It also enabled identifying and addressing the shortcomings of the generative models used for embodied AI.
\end{abstract}
\section{Introduction}\label{sec:intro}

Since the advent of Large Language Models (LLMs), there has been a growing interest in their applications in embodied AI agents~\cite{huang2022}~\cite{reed2022}.
This effort has accelerated with the introduction of tool or function calling mechanisms, leading to the creation of more complex embodied AI systems~\cite{wang2023}.
These systems often rely heavily on custom implementations that enable integration with foreign APIs and AI models. 

Some existing solutions facilitate the development of tools for API integration, such as ModelScope~\cite{li2023}.
Systems such as ROSA~\cite{royce2024} are tightly coupled with particular solutions (in this case ROS and ROS~2\cite{ros2}), but do not offer solutions for integration with other communication stacks.
Frameworks such as AgentGYM~\cite{xi2024} offer agents a standardized environment and evaluate LLM-based agents within that environment.
These LLM-based solutions are generally designed for single-agent deployments.
However, all of the above inherit common limitations of LLMs, including hallucinations~\cite{huang2025}, a lack of interpretability and explainability~\cite{zhao2024}, and susceptibility to jailbreak attacks~\cite{xu2024}.
These constraints often lead to serious functional and ethical concerns.

It is worth mentioning that before the rise in popularity of LLMs, several frameworks were already developed for building robotic agents.
For example, FABRIC~\cite{seredynski2019} has been introduced to develop robotic agents that utilize reusable components for seamless integration with ROS.
Likewise, MARC~\cite{duan2023} has been designed for Multi-Agent Systems (MAS) based on reinforcement learning.
In addition to these, some frameworks have been proposed for specific robotic applications~\cite{zhu2006,lim2009,rogers2006}, ranging from designs based on multi-robot cooperation to human-robot interaction with multiple people.
An important contribution toward creating a generalized framework applicable across various robotic domains is the G++ architecture~\cite{cubillos2010}.
All of the aforementioned approaches lack the flexibility and potential for human-robot interaction offered by native integration with LLMs.

In this context, this paper presents RAI, a framework designed for creating Multi-Agent Systems (MAS) with built-in capabilities for integration with LLMs, robotic stacks, external APIs, and human-robot interaction mechanisms. 
It also comes with components which support embodiment, such as tools for understanding Agent's physical form.
Generality is achieved through a modular and scalable architecture, which is based on the M-Agent model~\cite{cetnarowicz1996}.
This model defines an agent through its reasoning system, sensors, and actuators.

In RAI, these abstractions are streamlined into the agent's core logic and connectors -- utility components that allow the agent to both observe and interact with a broadly defined environment.
RAI agents can be virtual (interacting solely with a digital environment), physical, or exist within a virtual-physical continuum, seamlessly integrating both domains.
By being in a single MAS, the virtual and physical agents can communicate with each other to better achieve their goals.
The details of these mechanisms are described in Section~\ref{sec:arch}.

This paper presents three deployments in different environments to validate the viability of RAI.
The performance of the system in an edge computing setup within a physical environment has been tested using ROSBotXL, a ROS~2 autonomous mobile robot platform. 
RAI agents have been deployed on the platform to observe the environment, control robot movement, and enable human-robot interaction (HRI) capabilities.
Details of this experiment are presented in Section~\ref{ssec:rosbot}.
In addition, two pure virtual (simulation-based) experiments have been performed.
These RAI agents have been utilized for manipulation tasks with a simulated robotic arm and controlling a tractor.
The details are described in Section~\ref{sec:deploy}.

\section{Architecture}\label{sec:arch} % Kajetan
RAI architecture defines high level abstract classes, that allow for concise and flexible Agent definition.
The current section focuses on details of these interfaces and describes example use cases.
\subsection{Component definitions}\label{ssec:overview}

At the highest level, RAI uses three abstractions -- \textit{Agents}, \textit{Connectors}, and \textit{Tools}.\newline
\textit{\textbf{Agents}} are the basic building blocks of a MAS.
RAI \textit{Agent} implements a simple interface containing two methods, \texttt{run()} and \texttt{stop()}.
\textit{Agents} are also guaranteed to possess \textit{Connectors} -- mechanisms for observing and actuating their environment.
It is up to the user to define the internal model of the environment and the decision-making process for selecting actions.
Some RAI \textit{Agents} use simple rules-based systems, while others can utilize mechanisms such as LLMs for the decision-making process.\newline
\textbf{\textit{Connectors}} are an abstraction that represents the sensors and actuators of the \textit{Agents}.
They enable three modes of communication: publish-subscribe, service-based, and action-based.
Previous experiences~\cite{holda2022} have shown that relying only on one mode of communication can be limiting, so with RAI, we decided on an approach similar to ROS~2 communication.
The publish-subscribe mode provides utilities for publishing and receiving messages from other hosts.
In most implementations (e.g., ROS~2, MQTT, XMPP), this method of communication is efficient and performant. However, its reliability is limited as there is no guarantee that a sent message will be received.
The service-based mode addresses this limitation by providing the API caller with information on the message delivery and the result of the call.
This type of communication is usually less performant than the publish-subscribe mode (given the need to receive a response), but provides more reliability.
Finally, the action-based mode is modeled after the ROS~2 Actions API, which informs users about the acceptance and outcome of actions.
In addition, it provides continuous feedback on the execution state.

A specific RAI \textit{Connector} can implement all of these modes or combine any of them depending on the particular communication mechanism it encapsulates.\newline
\textbf{\textit{Tools}} are an abstraction inspired by LLMs and their tool-calling mechanisms.
A specific \textit{Tool} implements a way to create or parse data sent or received through a \textit{Connector}.
These implementations are compatible with langchain~\footnote{\url{https://www.langchain.com/}}, enabling seamless integration with tool-calling-enabled LLMs.
They can also be used by \textit{Agents} utilizing other decision-making mechanisms.

\subsection{Embodiment}\label{ssec:embodiment}

Robotic embodiment lacks a clear and broadly accepted definition.
Some researchers~\cite{Guzman2022} argue that simply having information on ways to interact with an environment is sufficient to constitute embodiment.
Others~\cite{duffy2000} think that more information is needed, such as a certain understanding of an environment and a robot's own ``body''.
RAI follows the latter view by enabling the creation of an agent's embodied identity from versatile data sources.
The proposed RAI framework enables users to experiment with different types and configurations of embodiment mechanisms.
The dedicated \texttt{RAI\_whoami} package provides an easy-to-configure way to incorporate embodying information into the \textit{Agents}.
This package includes features to automatically convert text, PDFs, and other documents into a vector database.
RAI supports multiple vector databases and was deployed with Faiss~\cite{douze2024faiss} as a robust solution.
The vector database can then enable LLMs with Retrieval Augmented Generation (RAG).
The functionality to store other embodiment-related data, such as images or URDFs, is also provided.
All of this data is accessible to a RAI \textit{Agent} and can be used to semi-automate the creation of system prompts.
It can also be used during the \textit{Agent's} runtime to retrieve additional information, e.g. when handling a user requesting information about the robot's capabilities.

%B: I think we should add some section about embodiment/docs parsing, RAG. Images - system prompt.

\subsection{Example of system configuration}\label{ssec:usageexamples}
% P: "Example of system configuration"?

\begin{figure}[htpb]
    \centering
    \includegraphics[width=0.9\linewidth]{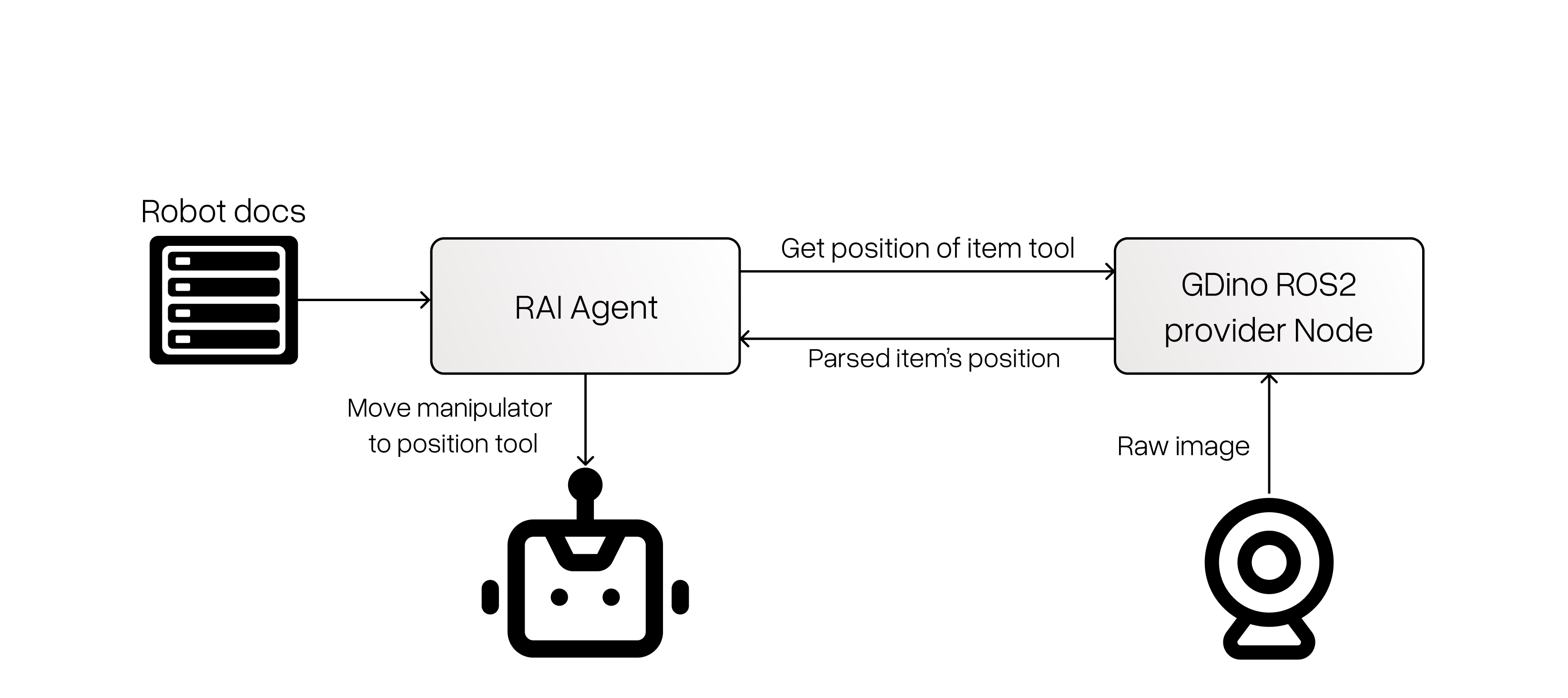}
    \caption{Simple embodied agent setup}
    \label{fig:gdino}
\end{figure}

To better understand these abstractions, consider a simple embodied deployment illustrated in Figure~\ref{fig:gdino}.
In the illustrated example, we have access to a robot, its documentation, and a ROS~2 service that uses an open-set segmentation model GroundingDINO~\cite{liu2023}.
These can be easily connected to create an embodied \textit{Agent} using RAI.
Such an \textit{Agent} would be designed to have two ROS~2 \textit{Connectors} -- one for connection with the robot and one with the GroundingDINO ROS~2 Node.
The first \textit{Connector} would be coupled with \textit{Tools} designed for robot control, and the second one with a \textit{Tool} for parsing GroundingDINO output into LLM understandable input.
The \textit{Agent} could be a simple conversational LLM, or a more advanced system (see Section~\ref{sec:deploy} for more complex examples).

% P: If there will be some places left, I would make this image bigger. And I would consider making the font bigger too.
% \FloatBarrier
\subsection{Flexibility and extensibility}\label{ssec:library}
% P: I would also consider moving this section before 2.3 -> Architecture -> Embodying RAI -> Flexibility -> Example. Up to you, this is not a strong suggestion :)

\begin{figure}[htpb]
\centering
\begin{minipage}{.5\textwidth}
  \centering
  \includegraphics[width=0.8\linewidth]{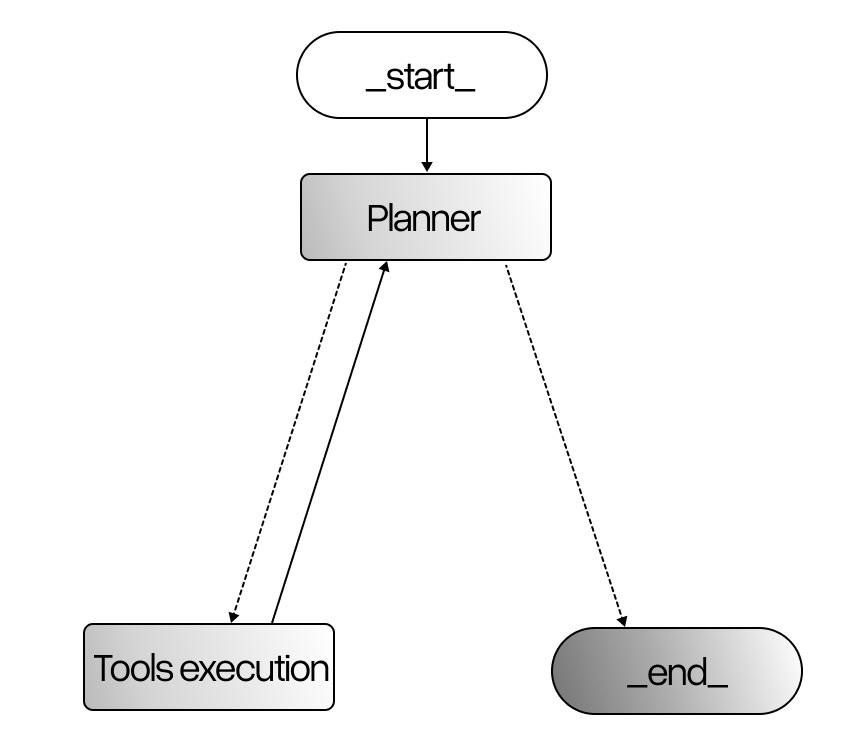}
  \captionof{figure}{Conversational Agent}
  \label{fig:conversational}
\end{minipage}%
\begin{minipage}{.5\textwidth}
  \centering
  \includegraphics[width=0.8\linewidth]{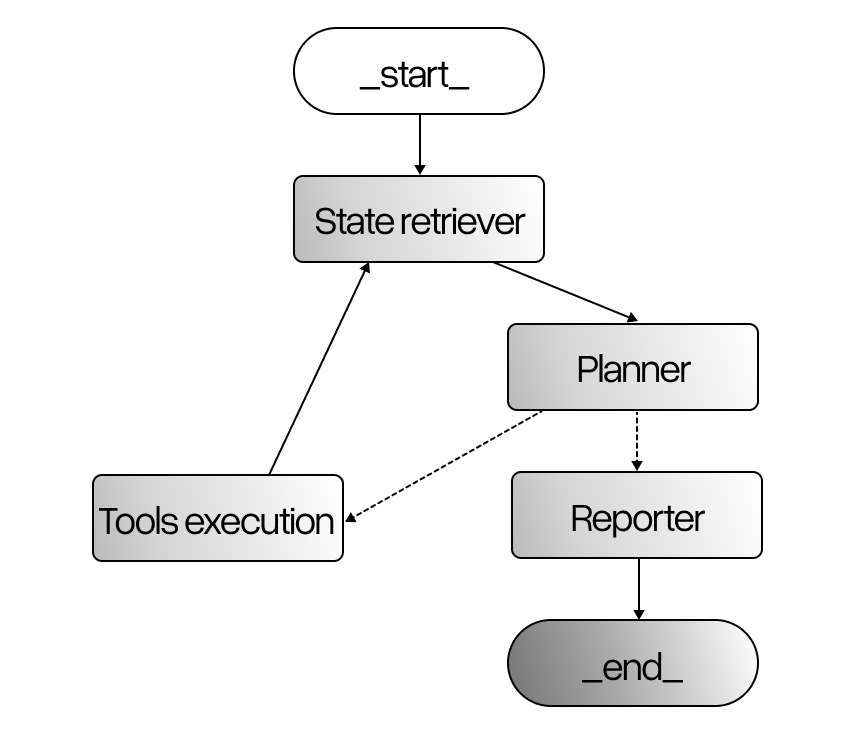}
  \captionof{figure}{StateBased Agent}
  \label{fig:statebased}
\end{minipage}
\end{figure}

To simplify the deployment of abstractions from Section~\ref{ssec:overview}, RAI comes with a library of configurable components.
Currently, there are four pre-packaged deployable agents:
\begin{itemize}
    \item \texttt{VoiceRecognition} -- coupled with configurable models for speech processing pipelines, enabling automatic speech recognition and transcription for human-robot interaction, as well as microphone integration;
    \item \texttt{TextToSpeech} -- equipped with direct speaker integration and coupled with text-to-speech (TTS) model;
    \item \texttt{Conversational} (Figure~\ref{fig:conversational}) -- an agent based on ReAct~\cite{yao2022react} LLM architecture which utilizes connectors to send and receive multimodal inputs;
    \item \texttt{StateBased} (Figure~\ref{fig:statebased}) -- an agent based on a finite state machine -- a mechanism similar to SPADE's~\cite{palanca2022} \textbf{FSMBehaviour}, which integrates LLM based reasoning with procedural approaches, which can be utilized for more complex tasks.
\end{itemize}

% P: Font is rather too small to read this comfortably.

The provided \textit{Agents} come with a set of connectors that enable interoperability through communication protocols (e.g. \texttt{ROS2Connector}) and direct peripheral access (e.g. \texttt{SoundDeviceConnector}).
Connectors are coupled with tools for low-level communication e.g. \texttt{CallROS2Service} or \texttt{ReceiveROS2Message}.
The library also includes \textit{Tools} for data processing, like \texttt{GetDistanceToObjects} for spatial analysis of camera and depth data.
It should be noted that the RAI components library is under active and continuous development, so the pool of available components is constantly growing.
In Section~\ref{sec:conclusions} plans for further work are described.

\section{Applications}\label{sec:deploy}
\subsection{Navigation and Human-Robot-Integration: Autonomous Mobile Robot}\label{ssec:rosbot}

\subsubsection{Overview}
\begin{figure}
\centering
\begin{minipage}{.5\textwidth}
  \centering
  \includegraphics[width=0.95\linewidth]{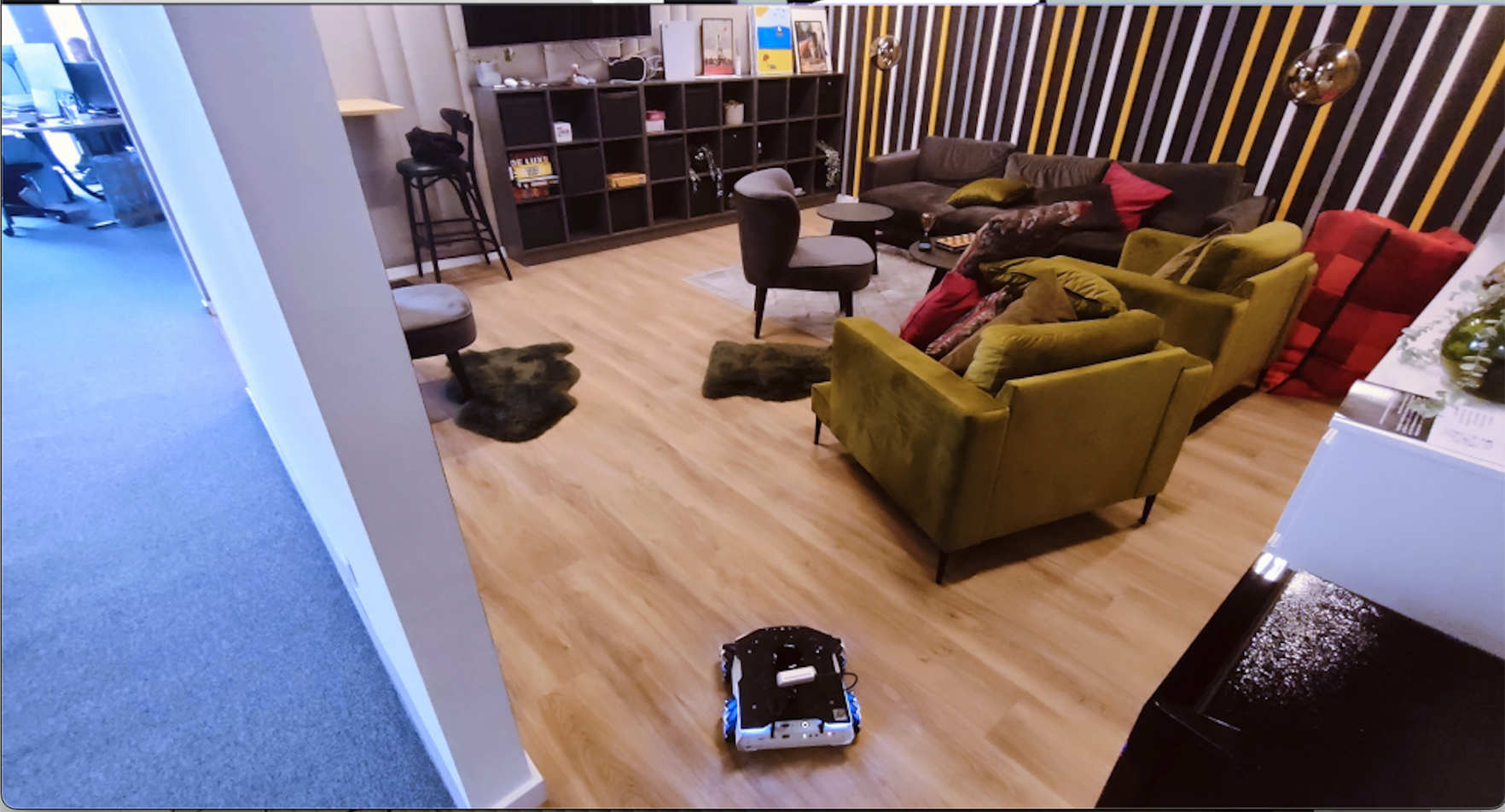}
  \captionof{figure}{Physical office setting}
  \label{fig:physical}
\end{minipage}%
\begin{minipage}{.5\textwidth}
  \centering
  \includegraphics[width=0.95\linewidth]{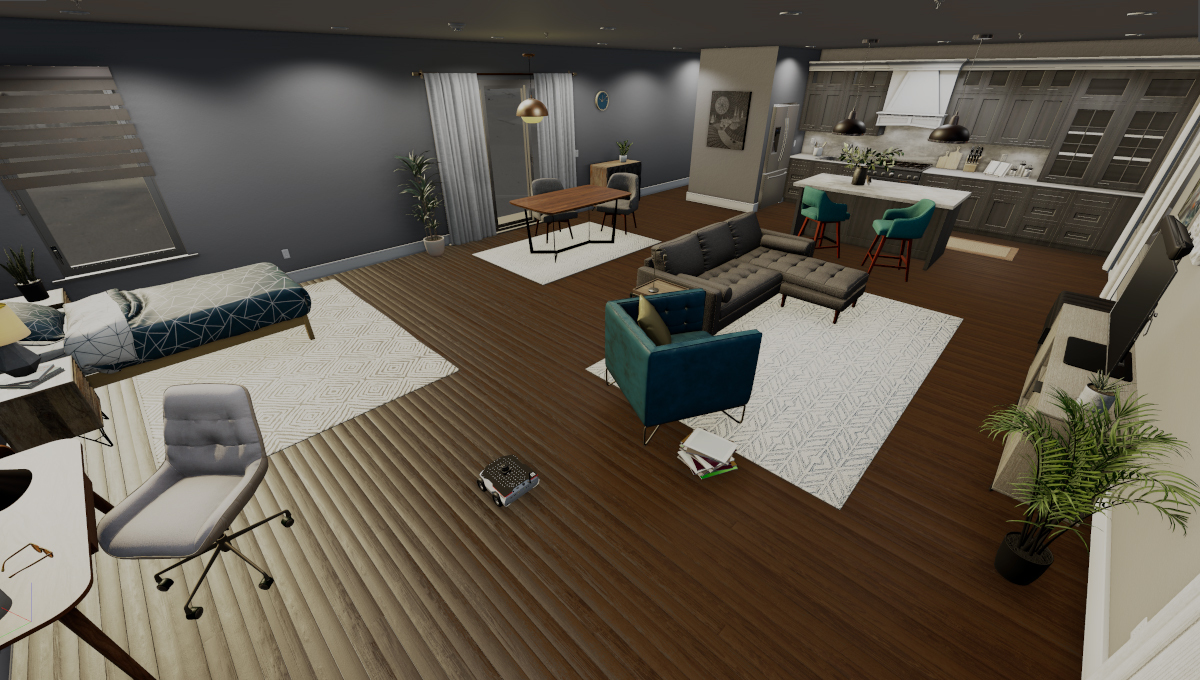}
  \captionof{figure}{Household with the digital twin}
  \label{fig:digitaltwin}
\end{minipage}
\end{figure}

The Husarion ROSBot XL was selected for indoor navigation and HRI experiments.
It is an autonomous mobile robot platform for indoor applications.
It comes equipped with an Intel RealSense Camera, a Slamtec RPLIDAR S3, and a Bosch BNO055 IMU.
Navigation is controlled through the ROS~2 nav2~\cite{macenski2020marathon2} stack.
Two environments were used for the tests: a physical office setting (Figure~\ref{fig:physical}) and a simulated household implemented in Open 3D Engine (O3DE)~\footnote{\url{https://o3de.org}} (Figure~\ref{fig:digitaltwin}) with a digital twin of the robot.
The double deployment accelerated prototyping and enabled testing of the system's robustness during the transfer from simulation to the real world.
In addition, the MAS deployment included a flexible human-robot interface that supports both S2S and text-based interactions via a web user interface.

In the experiments, RAI \textit{Agents} were equipped with open-set detection and generic ROS~2 tools used by the \textit{Agents} for the discovery of robot interfaces and the execution of navigation objectives.
To achieve embodiment, the system prompt was initialized with the robot's identity, and a tool for RAG querying was provided utilizing the mechanism described in Section~\ref{ssec:embodiment}.
% P: I would make these into a two point bullet list, just for readability (and because of sentence contructions).
\subsubsection{Configuration}
\begin{figure}[htpb]
    \centering
    \includegraphics[width=0.75\linewidth]{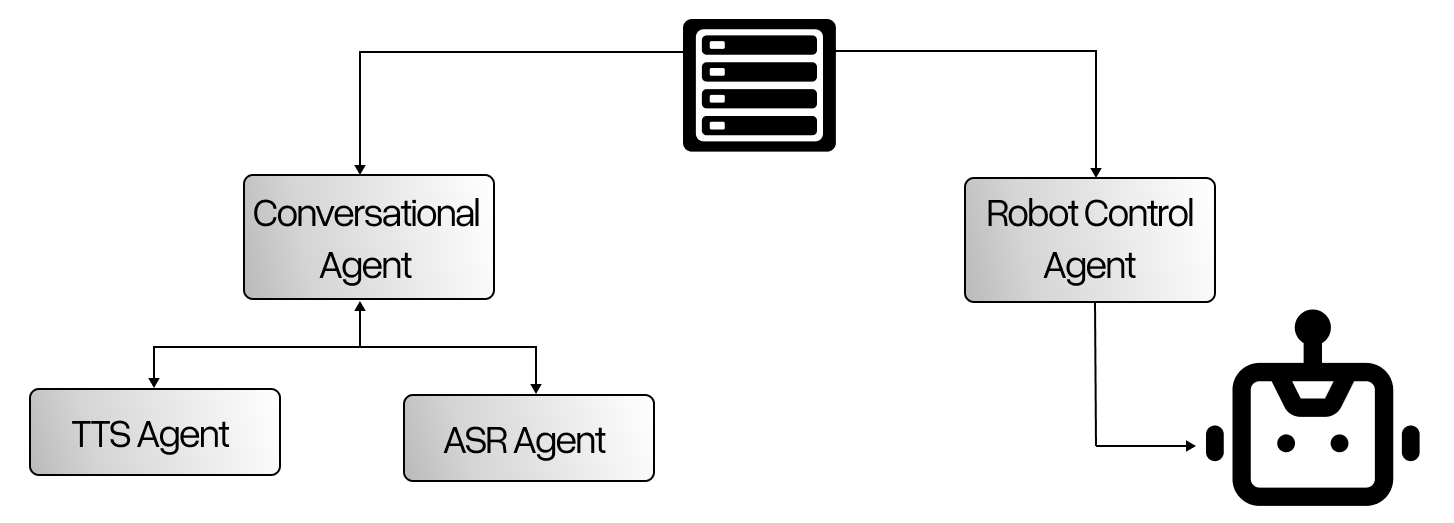} 
    \caption{Multi-agent architecture for S2S and navigation}
    \label{fig:rosbotarchi}
\end{figure}

A series of experiments was conducted using different architecture configurations.
Initially, a single state-based agent (described in Section~\ref{ssec:library}) was implemented to handle human interaction, camera image interpretation, and tool invocation.
This configuration resulted in advantages such as lower latency, full context retention, and easier debugging.
However, it also showed drawbacks: while the robot was performing a mission, the HRI interface was less responsive, as the LLM risked losing track of the mission goal when handling concurrent user queries.
To overcome these issues, a multi-agent setup was applied -- shown in Figure~\ref{fig:rosbotarchi}.
The Conversational \textit{Agent} was responsible for HRI and ensured continuous responsiveness to user input.
The Robot Control \textit{Agent} was focused solely on the execution of the mission.
The mission was defined by a human prompt (e.g. ``Navigate to the chair'').
Its execution was performed through ROS~2 based tooling using actions.
The \textit{Agent} was also responsible for deciding when to report completion or failure to the user.
\subsubsection{Conclusions}
The Robot Control \textit{Agent} successfully achieved navigation and object detection capabilities using the ROS~2 nav2 stack and the open-set detection tools.
The system was able to handle commands such as navigating to specified locations or positioning itself relative to environmental objects.
The digital twin environment proved invaluable for rapid prototyping, and the subsequent transfer to the real robot confirmed the robustness of the design.
In general, the MAS-based setup provided a balanced approach between efficient mission execution and effective human-robot interaction.
The framework proved itself to be flexible enough for testing different complex agent setups, where the user can examine the influence of each pipeline component.

\subsubsection{Challenges}
Several challenges were encountered during system integration.
Difficulties were experienced in achieving a clear understanding of agent embodiment, and issues were observed in the multi-agent architecture. 
In particular, inconsistencies in the understanding of embodiment were occasionally observed between the two LLM-based \textit{Agents}.
It has also been noted, that the decision to make LLM-based \textit{Agents} responsible for synchronization of information regarding the state of the mission led to some inconsistencies.
Similar problems were also encountered with long-running background tasks executed by the robotic stack. 
This could be addressed by incorporating rule-based synchronization mechanisms in future work.
Furthermore, while Robot Control \textit{Agent} could use the navigation stack successfully, error handling and mission success detection were inconsistent. 

\subsection{Manipulation: Robotic arm}\label{ssec:arm}

\subsubsection{Overview}
An \textit{Agent} controlling a robotic arm was deployed in O3DE, where it received a task defined using natural language. 
\textit{Agent's} perception was limited to a camera stream.
The received images were processed using an open-set segmentation model~\cite{ren2024grounded}, which provided additional information about the view.
The observed environment included a desk with colorful blocks and vegetables.
The interaction with the environment was performed with \textit{Tools} utilizing MoveIt 2~\cite{DBLP:journals/corr/ColemanSCC14} for motion planning and execution.

\subsubsection{Experiments}
\begin{figure}[htp]
    \centering
    \includegraphics[width=.75\linewidth]{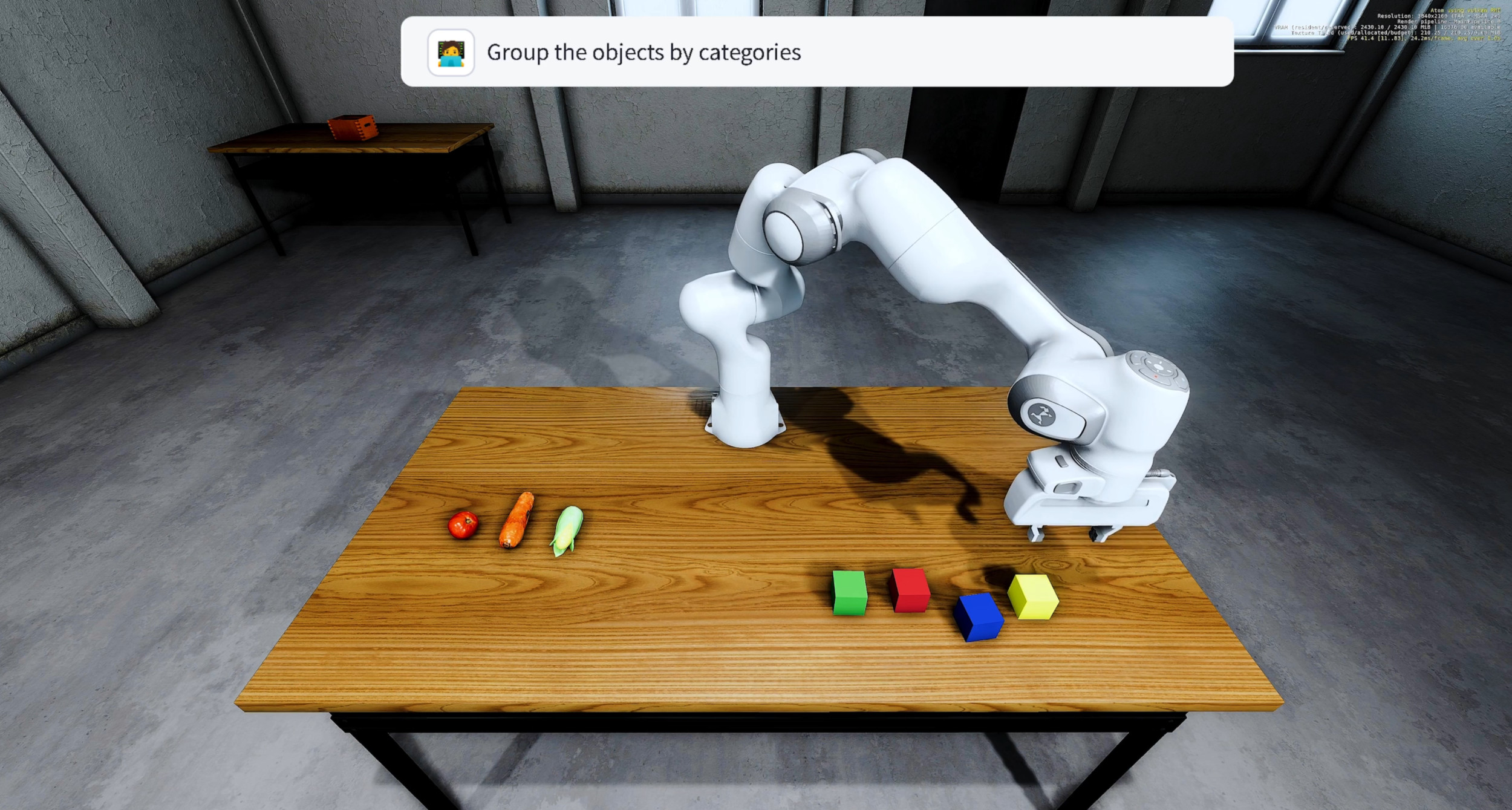}
    \caption{Scene setup after a sorting task is completed}
    \label{fig:arm_img}
\end{figure}

To evaluate its performance, the \textit{Agent} was tested in three key manipulation tasks:
\begin{enumerate}
    \item Sorting Objects – classifying objects before placing them in corresponding groups (Figure~\ref{fig:arm_img});
    \item Stacking Items – building stable stacks by placing objects on each other;
    \item Object Replacement – swapping pairs of objects, strategically using an intermediate position to prevent collisions and ensure proper placement.
\end{enumerate}
Various cloud-based multimodal LLMs (GPT-4o, GPT-4o mini, Claude 3.5 Sonnet) were tested as reasoning engines.
They were tasked with the interpretation of scene data, planning movement sequences, and verifying task success.

\subsubsection{Challenges}
Although success has often been observed in these tasks, \textit{Agents} repeatedly faced the same difficulties, particularly in spatial reasoning and self-correction.
When stacking (task 2), the \textit{Agent} often tried to place objects ``inside'' each other rather than on top, failing to account for physical boundaries.
A similar issue occurred during swapping objects (task 3) -- the \textit{Agent} tried to place the moved object directly in the place of the second object, without considering an intermediate position.
Despite explicit instructions to verify the completion using camera images, the \textit{Agent} frequently misjudged its success, assuming that the tasks were completed correctly even when errors were present.
These problems were observed in multiple LLMs, suggesting general limitations in spatial reasoning, understanding boundaries, and object interactions.

\subsubsection{Results}
The \textit{Agent} succeeded in basic manipulation tasks.
However, it frequently failed on tasks that require spatial reasoning and action sequencing.
Stacking and object replacement errors were common, as the \textit{Agent} lacked an intuitive grasp of the physical constraints.
The \textit{Agent} frequently failed to recognize its mistakes due to poor image understanding within the simulation.
Most of the LLMs tested were able to complete the assigned tasks with adequate prompt engineering.
However, in many cases, the instructions had to be extremely explicit, essentially guiding the LLM step-by-step through each action.
Without such detailed instructions, the \textit{Agent} frequently struggled to perform an effective sequence of movements.
This highlights a key limitation: while multimodal LLMs can perform complex tasks, their ability to independently reason about spatial interactions remains limited.

\subsection{Agriculture: Handling edge-cases}\label{ssec:agri}
\begin{figure}[htpb]
    \centering
    \includegraphics[width=0.55\linewidth]{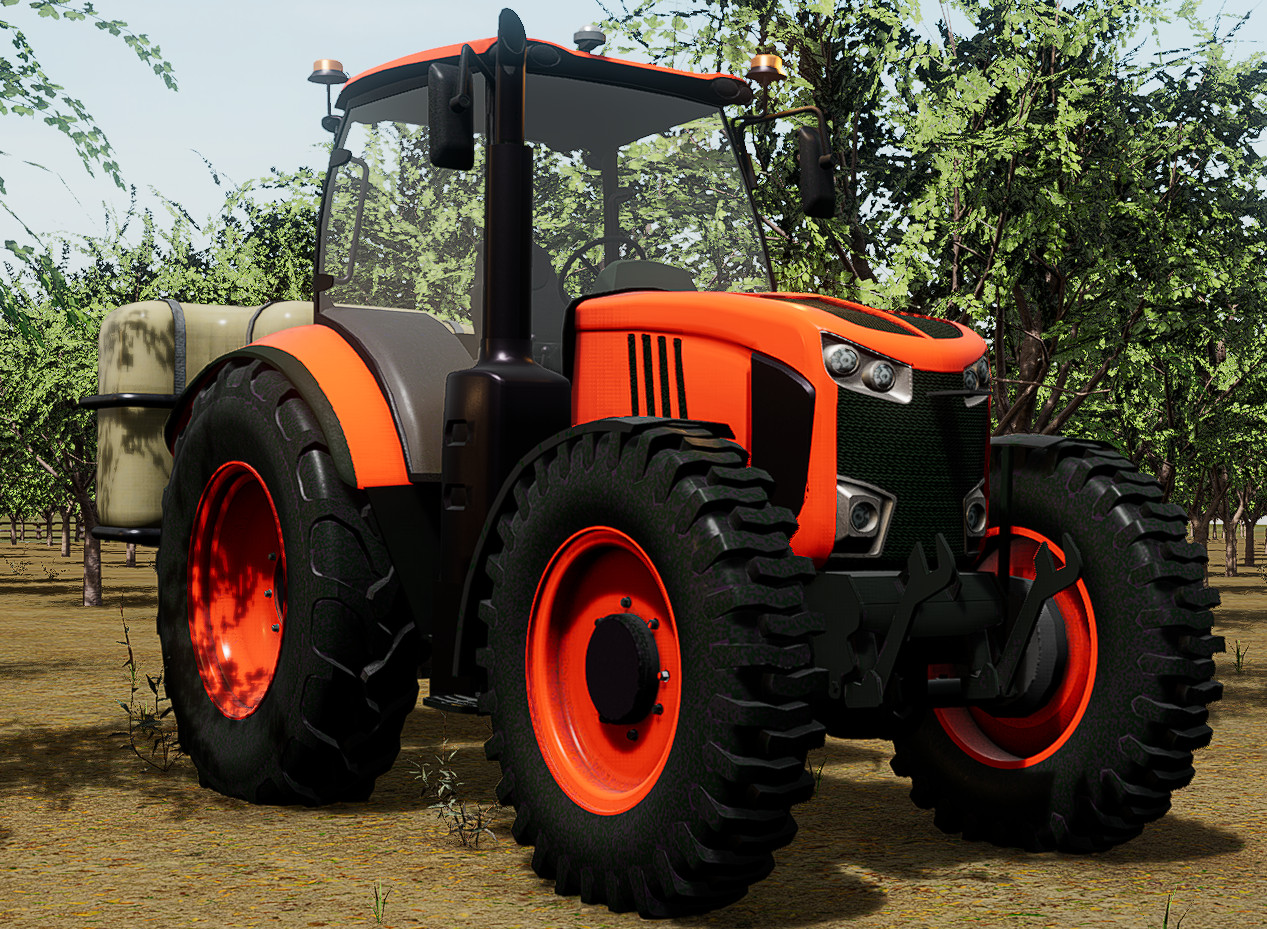}
    \caption{Tractor's image used for enhancing Agent's embodiment}
    \label{fig:agri_img}
\end{figure}
\subsubsection{Overview}
An autonomous tractor was deployed in O3DE, a ROS~2-enabled simulation environment.
The environment was designed for high-fidelity real-time simulation of an orchard with trees. 
The \textit{Agent} was specifically tasked with handling unexpected situations in an agricultural setting.

The tractor was equipped with a rule-based autonomous system.
When that system detected an anomaly, the tractor's control was transferred to the \textit{Agent}.
At that point, \textit{Agent} gathered environment understanding through a front camera image and had to determine the most appropriate response to ensure both safety and efficiency. 
The available \textit{Tools} included replanning the route, driving forward (over the obstacle), using visual and auditory signals, or aborting the task.

\subsubsection{Experiments}
To evaluate the \textit{Agent}'s performance in agricultural automation, two experimental conditions were tested:

\begin{enumerate}
    \item Language-only embodiment: a natural language description of the tractor, specifying its size, movement capabilities, and operational constraints;
    \item Visual embodiment: an image (Figure~\ref{fig:agri_img}) of the tractor, in addition to the description, offering it a direct visual reference of its physical attributes. 
\end{enumerate}
A range of multimodal LLMs (GPT-4o, GPT-4o mini, Claude 3.5 Sonnet) was tested as reasoning engines for the agent.
The experiments focused on response to anomalies, specifically recognizing, classifying, and responding to obstacles in the environment.

\subsubsection{Challenges}

The primary issue with the language-only embodiment was misjudging the durability of the tractor.
Without a visual reference, the \textit{Agent} was overly cautious, often stopping for small obstacles such as branches, incorrectly assuming they could cause damage.
% This caused unnecessary interruptions and inefficiencies in the execution of tasks.

The visually embodied \textit{Agent} performed significantly better in assessing the tractor's physical capabilities.
Both \textit{Agents} occasionally misidentified objects, leading to errors in decision-making.

\subsubsection{Results}
Providing the \textit{Agent} with a visual reference of the tractor improved its obstacle assessment.
With an image of itself, the \textit{Agent} had a better grasp of its dimensions and durability.
Unlike the language-only \textit{Agent}, it correctly identified that small branches did not pose a threat.
The setup including vision-based embodiment led to an increase in correct hazard classification.

Despite these improvements, both agents struggled with image-based object recognition.
Occasionally, classification errors led to incorrect hazard assessments, unnecessary maneuvering, or aborting the mission.
While the visual embodiment helped mitigate false positives, the underlying image processing remained a limitation.

\section{Conclusions and Future Work}\label{sec:conclusions}% Kajetan

The goal of this work was to develop a scalable, extensible, and flexible system to implement embodied MAS in both simulated and physical environments.
To achieve this, the RAI framework was introduced, designed to address common challenges in working with autonomous and semi-autonomous robotic agents.

The RAI’s architecture supports scalability by allowing the addition of new \textit{Agents} with minimal overhead.
It is extensible, enabling users to develop custom \textit{Tools} and functionalities, and flexible, simplifying integration across various deployment scenarios.
These capabilities have been demonstrated through successful deployments on multiple robotic platforms, highlighting RAI’s potential for embodied agent applications in diverse contexts, as demonstrated in Section~\ref{sec:deploy}.
The framework has also proven valuable for experimenting with embodiment strategies and evaluating LLM capabilities.

The current version of RAI is available at \url{https://github.com/RobotecAI/rai}.
It should be noted that the framework undergoes constant development.
Further work will include expanding the components library (Section~\ref{ssec:library}) and adding features meant to address limitations of LLMs (e.g. spatio-temporal database, or knowledge streaming).
Contributions to its further development are welcome and requested.

% \begin{credits}
% \subsubsection{\ackname} A bold run-in heading in small font size at the end of the paper is
% used for general acknowledgments, for example: This study was funded
% \subsubsection{\discintname}
% It is now necessary to declare any competing interests or to specifically
% state that the authors have no competing interests. Please place the
% statement with a bold run-in heading in small font size beneath the
% (optional) acknowledgments\footnote{If EquinOCS, our proceedings submission
% system, is used, then the disclaimer can be provided directly in the system.},
% for example: The authors have no competing interests to declare that are
% relevant to the content of this article. Or: Author A has received research
% grants from Company W. Author B has received a speaker honorarium from
% Company X and owns stock in Company Y. Author C is a member of committee Z.
% \end{credits}
\bibliographystyle{splncs04}
\bibliography{refs}
% LNCS Homepage, \url{http://www.springer.com/lncs}, last accessed 2023/10/25
% \end{thebibliography}
\end{document}